# A review-based study on different Text-to-Speech technologies

Md. Jalal Uddin Chowdhury, Ashab Hussan
Leading University, jalalchy101, ashabhtanim@gmail.com

*Abstract* - **This research paper presents a comprehensive review-based study on various Text-to-Speech (TTS) technologies. TTS technology is an important aspect of human-computer interaction, enabling machines to convert written text into audible speech. The paper examines the different TTS technologies available, including concatenative TTS, formant synthesis TTS, and statistical parametric TTS. The study focuses on comparing the advantages and limitations of these technologies in terms of their naturalness of voice, the level of complexity of the system, and their suitability for different applications. In addition, the paper explores the latest advancements in TTS technology, including neural TTS and hybrid TTS. The findings of this research will provide valuable insights for researchers, developers, and users who want to understand the different TTS technologies and their suitability for specific applications.**

*Index Terms* – Natural Language Processing, Text-to-Speech.

## INTRODUCTION

Nowadays, we are living in the digital era. Every staff associated with our life is getting digital. Almost every smartphone has a smart assistant that can speak and communicate like a human. Speech recognition is one of the technologies used in those smart assistants. Text-to-Speech is a part of speech recognition. Text-to-speech (TTS) is a natural language modeling approach that converts text units into speech units for audio presentation. There are numerous technologies used in Text-to-Speech. Many programming languages are used in these technologies. Python, a programming language mostly used in Text-to-Speech technology. There are many Python library e.g gTTS, pyttsx3, paddlespeech. But everyone's performance is not the same. Our thesis is about measuring the efficiency of various Text-to-Speech technology in various aspects.

*I. How to Works TTS*

Text-to-speech converts text into human-like speech, along with the ability to create a unique, custom voice.

### A. Type of Voice:

**Standard voice :** Standard voice is the simplest and most cost-effective type of voice. In the past few years standard voice has improved considerably to provide a human-like voice in multiple regional dialects, such as Hindi or Irish English. Regional dialects provide greater clarity of pronunciation of region-specific words or phrases, making for more understandable and accessible accents.

**Neural Voice:** Neural voice is a new type of synthesized speech that's nearly indistinguishable from human recordings. Powered by deep neural networks, neural voices sound more natural than standard voices by producing human-like speech patterns, such as stress and loudness of individual words. Because of this human-like speech, you end up with a more precise articulation of words, along with a significant reduction in listening fatigue when users interact with AI systems.

**Custom Neural Voice:** Custom neural voice uses your own audio data to create a one-of-a-kind customized synthetic voice. Custom neural voice has the deepest level of voice personalization, with realistic speech that can be used to represent brands, personify machines, and allow users to interact with applications conversationally.

### B. Some Terminology

**Phoneme:** A phoneme is the smallest unit of sound that makes a word's pronunciation and meaning different from another word.

**Prosody:** The patterns of rhythm and sound used in poetry.

**Mel-spectrogram:** It is derived by applying a non-linear transformation to the frequency axis of short-time Fourier transform (STFT) of audio, to reduce the dimensionality. It emphasizes details in low frequencies which are very important to distinguish speech and de-emphasizes details in high frequencies which usually are noise.

**Text-To-Speech (TTS) Structure**

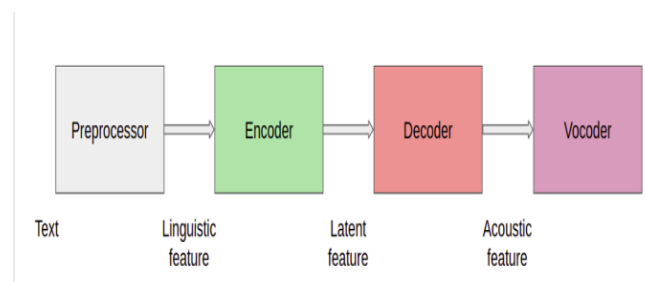

Fig. 1 : Text-to-speech structure

This is a high-level diagram of different components used in the TTS system. The input to our model is text, which passes through several blocks and eventually is converted to audio.

### Preprocessor

- **Tokenize**: Tokenize a sentence into words
- **Phonemes/Pronunciation:** It breaks input text into phonemes, based on their pronunciation. For example, "Hello, Have a good day" converts to HH AH0 L OW1, HH AE1 V AH0 G UH1 D D EY1.
- **Phoneme duration:** Represents the total time taken by each phoneme in the audio.
- **Pitch**: Key feature to convey emotions, it greatly affects the speech prosody.
- **Energy**: Indicates frame-level magnitude of mel-spectrograms and directly affects the volume and prosody of speech.

The Linguistic feature only contains phonemes. Energy, pitch, and duration are actually used to train the energy predictor, the pitch predictor, and the duration predictor respectively which are used by the model to get a more natural output.

### Encoder

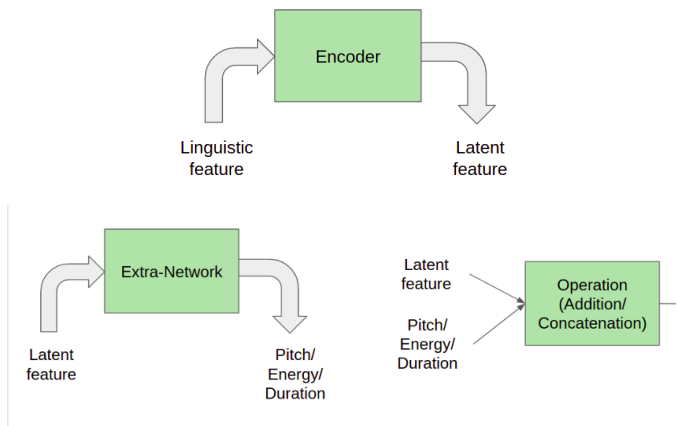

The encoder inputs Linguistic features (Phonemes) and outputs an n-dimensional embedding. This embedding between the encoder and decoder is known as the latent feature. Latent features are crucial because, other features like speaker embedding are concatenated with these and passed to the decoder. Furthermore, the latent features are also used for the prediction of energy, pitch, and duration which in turn play a crucial role in controlling the naturalness of the audio.

### Decoder

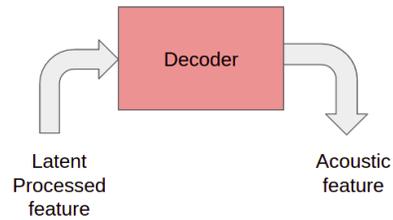

The decoder is used to convert information embedded in the Latent processed feature to the Acoustic feature i.e. Mel-spectrogram.

### Vocoder

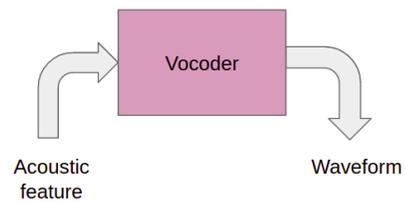

It converts the Acoustic feature (Mel-spectrogram) to waveform output (audio). It can be done using a mathematical model like Griffin Lim or we can also train a neural network to learn the mapping from mel-spectrogram to waveforms. In reality, learning-based methods usually outperform the Griffin Lim method.

So instead of directly predicting waveform using the decoder, we split this complex and sophisticated task into two stages, first predicting mel-spectrogram from Latent processed features and then generating audio using mel-spectrogram.

## LITERATURE REVIEW

Designing an effective text-to-speech synthesis system is quite difficult. Building a whole TTS system requires completing several steps, including normalizing text, converting text to phonemes, identifying prosodic emotional content, and generating speech.

Speech synthesis for different languages has already been the subject of many research proposals. Before electronic signal processing was invented, some early scientists tried to make machines that could mimic human speech.

A Unit Selection approach for the text-to-speech synthesis using Syllabic was presented in [1]. In this paper, They select syllables as their unit- hence this was the first syllable based Text to Speech conversion system for Bangla language. In this System, It is necessary to conduct a substantial amount of testing with an even larger text corpus than the one they utilized as an experimental text corpus.

The research by F. Alam and colleagues resulted in the development of a speech synthesizer for the Bangla language[2,3]. The diphone concatenation method was used to make this system. It needs a dictionary that tells it how to say words so that it can talk. There are 93000 entries in the dictionary [3]. The proposed system makes voice data for a festival and adds support for the Bangla language to the festival using its embedded scheme scripting interface. It turns Bangla Unicode text into ASCII text based on the Bangla phone set. But there is no explanation of how the process of transliteration works. Also, there is no information about how the letter-to-sound (LTS) rules for words that aren't in the lexicon were made.

In [4], the author showed how a Bangla Text-to-Speech (TTS) system was designed and built from the raw level without using third-party speech synthesis tools. For building the system, they were looked at from two different angles: one based on phonemes and the other on syllables.This study was conducted on a very raw level, and the researchers used recordings of their own voices to produce phonemes and syllables. The syllable-based method showed higher quality speech than the phoneme-based method. But, limited syllable and phoneme data were used for the development process.

In [5], the author used a concatenative synthesis technique to make the system's speech sound natural.In this paper, They proposed a system that converted Bangla text to Romanized text based on Bangla graphemes set and by developing a bunch of romanization rules.They used the MBROLA diphone database and did not develop their own database.Also, The sound quality is not particularly natural in its presentation.

In [6], they present FastPitch, a fully-parallel text-to-speech model based on FastSpeech, conditioned on fundamental frequency contours. Pitch contours are predicted by the model during inference. The generated speech can be made more expressive, better match the meaning of the utterance, and ultimately more interesting to the audience by changing these predictions.

In [7], the authors of the paper propose LightSpeech, which leverages neural architecture search (NAS) to automatically design more lightweight and efficient models based on FastSpeech. They meticulously developed a fresh search space that includes a variety of lightweight and potentially efficient architectures after thoroughly profiling the components of the current FastSpeech model. Then, within the search space, NAS is used to automatically find well-performing architectures. The model developed by their method, according to experiments, achieved a 15x model compression ratio and a 6.5x inference speedup on the CPU while maintaining a comparable voice quality.

In [8], the authors made a rule-based system for normalizing Bangla text instead of a decision tree and a decision list for ambiguous tokens.In this paper, a lexical analyzer was developed to tokenize each NSW(Non Standard Words) by using a regular expression and the tool JFlex[9]. This was done based on semiotic classes. The main thing that the work was that it was done in sequences of tokenization, token classification, token sense disambiguation, and standard word generation. This work will be useful in the future because it combines TTS and Speech Recognition and compares the ways that rule-based systems and other classification systems handle ambiguity.

In[10],The authors developed an audio programming tool for blind and vision-impaired people to learn programming that is based on text-to-speech technology. In this paper, they demonstrate how users who use the tool can edit programs, compile, debug, and run them. The authors mentioned that these levels can all be voiced. As a programming language, they use C# for evaluation, and VisualStudio.NET is used to create the tool. Evaluations have demonstrated that the programming tool can support the implementation of software applications by blind and vision-impaired individuals and the achievement of equality of access and opportunity in information technology education. To communicate with a computer, vision-impaired people liked to use mouse events and blind people liked to use keyboards with shortcut keys defined in JAWS. This means there's not any inbuilt or intuitive systematic approach to handle the interaction with computers.

A diphone-based concatenative technique was utilized by the authors in the development of a speech synthesizer for the Bangla language [11].In addition to this unique collection of words, the tokenization of null-modified characters has been presented in this study. This is an important and, to put it mildly, a tough task for a text-to-speech program (TTS).

From the perception of the authors, despite the fact that over 1.6 billion Muslims live in the world and that Arabic is spoken by millions of people as an official language in 24 different nations, it has received less attention than other languages [12]. These considerations highlight the necessity, from the point of view of the authors, for an Arabic TTS that could be of the highest quality, be lightweight, and be absolutely free. A rule-based system with an exception dictionary for words that don't follow the letter-to-phoneme rules might be a much more sensible approach since the vowelized written text of Arabic bears the pronunciation rules with few exceptions from their perspective. This study developed a rule-based text-to-speech hybrid synthesis system that combined formant and concatenation approaches to produce speech that sounds natural enough. But for the lack of significant stressed syllables and intonation, the overall system might not perform intuitively as well as handle the differentiation in different arabic accents.

## CONCLUSION AND FUTURE DIRECTION

This review-based study has examined different Text-to-Speech (TTS) technologies and highlighted their advantages and limitations. The study has provided an overview of the basic functionalities of TTS systems and has shown how they have evolved over time, from rule-based systems to neural-based models. The study has also explored the impact of TTS on different industries, including education, entertainment, and healthcare. One of the key findings of this study is that the recent advancements in deep learning have significantly improved the quality of TTS systems. However, there are still several challenges that need to be addressed, such as the lack of emotional expressiveness and naturalness in synthesized speech, which can affect the user experience. In terms of future directions, further research is needed to improve the performance of TTS systems in terms of naturalness, expressiveness, and intonation. This can be achieved by developing more advanced algorithms that can capture the nuances of human speech and emotions. Additionally, more studies are needed to evaluate the effectiveness of TTS in various applications, such as language learning and speech therapy. Overall, TTS technology has the potential to revolutionize the way we communicate and interact with machines. As the technology continues to evolve, it will become increasingly important to address the limitations and challenges of TTS to ensure that it can be used to its full potential.